# Extended defects in natural diamonds: Atomic Force Microscopy investigation

Radmir V. Gainutdinov [1], Andrey A. Shiryaev[2,*], Yana Fedortchouk[3]

1) Institute of Crystallography RAS, Leninsky Pr. 59, 119071, Moscow, Russia
2) Institute of Physical Chemistry and Electrochemistry RAS, Leninsky Pr. 31, 119071, Moscow, Russia
3) Department of Earth Sciences, Dalhousie University, Halifax B3H 4R2, Canada

\* - corresponding author. Email: shiryaev@phyche.ac.ru OR a_shiryaev@mail.ru

**Abstract.** Surfaces of natural diamonds etched in high-pressure experiments in $H_2O$, $CO_2$ and $H_2O$-NaCl fluids were investigated using Atomic Force Microscopy. Partial dissolution of the crystals produced several types of surface features including the well-known trigons and hillocks and revealed several new types of defects. Besides well-known trigons and dissolution hillocks several new types of defects are observed. The most remarkable ones are assigned to anelastic twins of several types. The observation of abundant microtwins, ordering of hillocks and presence of defects presumably related to knots of branched dislocations suggests importance of post-growth deformation events on formation of diamond microstructure. This work confirms previous reports of ordering of extended defects in some deformed diamonds. In addition, the current work shows that natural diamonds deform not only by dislocation mechanism and slip, but also but mechanical twinning. The dominant mechanism should depend on pressure-temperature-stress conditions during diamond transport from the formation domain to the Earth surface.

## 1. Introduction

Investigation of defects in diamonds remains an active field of mineralogy and solid state physics for several decades. For Geosciences studies of the defects are important for reconstruction of growth and annealing conditions of natural diamonds. In addition, the extended defects provide unique information on conditions of post-growth deformation, which are difficult to address otherwise. In particular, investigation of deformation-related defects helps to reveal the deformation mechanism, which, in turn, is needed for understanding the PT-conditions of deformation. The studies of the deformation-related defects are important not only from the fundamental point of view, but also for gemology, due to the effect of defects on diamond colour.

The most commonly employed methods sensitive to extended defects in diamonds are Transmission Electron Microscopy (TEM) and Infra-red (IR) spectroscopy. Atomic Force Microscopy (AFM) is currently widely used for investigation of surface features of materials. However, its applications to diamond are not numerous. Usually AFM is employed for detailed examination of surfaces of CVD diamond films, possessing large relatively smooth surfaces (e.g., de Theije, 2001). Applications of AFM to natural diamonds is technically demanding due to small and fairly rough surfaces typical for etched stones and only few works employed this technique for investigation of the etch relief (Fedortchouk et al., 2011). In this work we report results of AFM investigation of defects in natural diamonds revealed by artificial etching. We show that AFM may provide considerable amount of new information not only about purely surface-related features but also reveal extended defects, present in the bulk of the diamond.

## 2. Experimental

Surface relief of several natural diamonds with sizes below 2 mm was studied using Scanning Probe Microscope Ntegra Prima (NT-MDT, Russia). In most cases the contact mode was employed, though tapping regime was also used occasionally. Employed silicon cantilevers (Micromasch, Estonia) are characterized by the following parameters: resonance frequency f~60 kHz, tip radius R~10 nm, stiffness constant k ~ 1 N/m. Several cantilever geometries were used. The microscope is installed in clean room TrackPore Room'05 with controlled atmosphere. The humidity in the room can be varied between 30-70 % and drift is less than ± 1% per hour. The temperature was in the range 25±5 °C and the drift was ±0.05 °C per hour. Such conditions allowed high reproducibility of obtained data and drastically minimize artifacts. The analysed areas on the diamond faces close to 111 were selected using optical microscope. Analysis of images was performed using Nova and in PA software.

The main body of presented results was obtained on four samples of natural kimberlitic diamonds artificially etched at 1 GPa and 800°C - 1350°C in a piston-cylinder apparatus (Fedortchouk et al., 2007, Hilchie and Fedortchouk, 2009). One of them (termed "NaCl") etched in $H_2O$-NaCl fluid preserved octahedral shape due to low weight loss; the other samples etched in $H_2O$-fluid (M45 and M56) and in $CO_2$-fluid (M57) preserved only 50-60% of their initial weight. For experimental details see Fedortchouk et al. (2007).

## 3. Results
### 3.1. Hillocks

Hillocks of various shapes and sizes are the most prominent surface features of the studied diamonds, examples are shown on Figs. 1 and 2. Similar formations were extensively described in mineralogical literature (e.g. Orlov, 1963). In most cases the observed hillocks possess certain asymmetry, which is explained by variations in the etch rate of different crystallographic faces. The trajectory of the hillock vertex corresponds to the crystallographic direction with the slowest dissolution rate (for review see Sangwal, 1987 and references therein). We have made an attempt to use Atomic Force Microscopy to obtain information about mechanism of material removal from the surface of a hillock by close examination of topography of its sides. Results of such investigation show that the finest resolvable vertical steps on the hillock size are close to 0.4 nm, which is close to the diamond lattice parameter (0.354 nm). Unfortunately, the exact crystallographic orientation of the hillocks slopes is difficult to determine unambiguously, but the obtained value of the step size suggests that it represents 2-4 monolayers at most. Interestingly, on some generally featureless regions of the 111 face the smallest resolvable vertical features (roughness) are characterized by similar values. Examination of featureless "flat" regions of the 111 face shows that their average roughness is between 0.25-0.3 nm. The 111 interlayer spacing for diamond is 0.206 nm. The obtained average roughness suggests that in the absence of extended defects the material removal proceeds on level of individual carbon atoms.

Examination of mutual arrangement of the hillocks on various samples and parts of the studied faces shows that in many cases the smallest hillocks and other small features are distributed not absolutely randomly: in most cases they form kind of bands (Fig. 2). One might consider this banding as an artifact resulting from convolution of the cantilever tip with the sample relief. However, this suggestion can be ruled out since completely independent experimental runs with different types of cantilevers and scans rates gave similar images. Weak periodicities may be established in some cases using Fourier analysis.

In addition to hillocks parallel bands of positive relief are also observed. In some cases these individual bands are made of microhillocks of rather similar sizes, forming rope-like features (Fig. 3). Hints for periodicity in direction perpendicular to bands direction are often observed.

Interestingly, hints for long range ordering of extended defects in diamonds is given by Small-angle scattering (Shiryaev 2007, Shiryaev and Boesecke, 2012) and by electron microscopy (Bangert, pers. comm). The ordering of extended defects is common phenomenon for ageing alloys and solid solutions, but in most cases it is observed when the concentration of impurity is in the order of several percents. In diamond such ordering might be related to interaction of stress fields from slip planes and other similar extended defects. Overlaps of the stress fields may promote quasi-periodic preferential materials removal during the etching. This model is supported by arrangement of deformation-related twins, see below. However, the observed regularity in hillocks distribution is likely unrelated to deformation (Lüders) bands (e.g. Likhachev et al., 1989).

## 3.2. Deformation-related features

One of the most remarkable features observed is represented by "ridges": positive forms of surface relief consisting of long (up to 10 microns), narrow (<2 mkm) features with heights usually below 40 nm (Figs. 4, 5). Two different types of the ridges were observed: a) blunted on one end (Fig. 4); and b) with both sharp ends (Fig. 5). Apparently, the type of the ridges is sample-dependent, though if our explanation of their origin is correct (see below), they may coexist in some diamonds. The spatial distribution of the ridges on faces of diamonds is very heterogeneous. The ridges are clearly not related genetically to other dissolution features such as trigons, hillocks etc., since in many cases they are superimposed on other dissolution-related features. This strongly suggests that the ridges represent defects in structure of individual stones.

Apparently, such ridges were not reported previously despite numerous studies of diamond surfaces using optical microscopy. The reason for this might be a combination of their small width and elevation, precluding easy observation. To the best of our knowledge the only SEM observation of such feature was reported by Khokhryakov & Pal'yanov, (2007) (fig. 1c of their paper) who have studied manifestations of extended defects in etched synthetic diamonds and have ascribed the feature to a stacking fault. These authors believed that the feature possessed *negative* relief. However, determination of the relief sign using SEM is not always trivial task and in reality the feature observed by these authors was, perhaps, protruding from the surface as is observed by us. At the same time, extensive works on etching of silicon have shown (e.g., Seiter 1977) that the composition of the etching agent, especially the oxidizer type and concentration, is very important for formation of the surface relief.

The most plausible explanation of the origin of these features is that they represent anelastic twins (see reviews by Kosevich & Boyko 1971, Boyko et al., 1994) and/or tent-shaped twins (Hirth et al., 1998). Such features were extensively studied on calcite crystals. The twin lamellae may appear on moderate loads and are often reversible, i.e. disappear when the load is released. However, in many cases they can be "frozen" in the sample and this is the case observed by us. The reason for that lies in restricted mobility of dislocations in solids with high Peierls barrier, such as diamond. Remarkably, the height of the observed ridges (≤40 nm) is similar to that observed by DeVries (1975) for deformation lamellae on polished diamonds.

Though the statistics is limited, Fourier analysis shows that on some crystals the twin lamellae are arranged in bands with weakly pronounced periodicity. This observation can be due to long-range stress field around the tent-shaped twins with eventual generation of an antitwin (Hirth et al., 1994) and may be of similar origin to preferential arrangement of hillocks mentioned above.

The twin lamellae are surrounded by twinning dislocations. If the dislocation mobility is low, as is in case of diamond at moderate temperatures, the twin may remain even when the load is removed. In our AFM images the twinning dislocations are manifested in two different ways: a) wavy boundary of the blunted twins (fig. 4 b-d) is likely due to etched dislocation cores and b) a very shallow depression around the twin in fig. 5. Interestingly, on the figure 5 one can see very early stages of appearance of a second twin. The difference between the blunted and sharp-ended twins most likely lies in microstructure of a particular diamond crystal. Theory of an evolving deformation twin shows that its advancing end should be sharp (Boyko et al., 1994). If a growing twin encounters an obstacle which is large enough, its leading end becomes rounded. Therefore, our work reveals two types of anelastic twins: those pinned at some defects and free twins.

In principle, the ratio of a twin width to its length is proportional to external load (Kosevich and Boyko, 1971). Unfortunately, in our case reliable determination of the load is barely possible, since the degree of the twin relaxation can not be estimated.

In the central lower and at the right part of the Figure 4b one notices an example of yet another twin cutting the 111 face at approximately 35 degrees: an invariant plane strain (ISP) type (e.g., Hirth et al., 1998). A closer look suggests that the defects observed on this view represent twin accommodation by slip in the matrix and in the twin itself (Christian and Mahajan, 1995). Apparently, their number density is markedly lower than that of anelastic twins, which is not unusual, since they may relax only by local slip.

Certain types of stacking faults, potentially present at twin boundaries, can be considered as lamellae of hexagonal diamond (h-dia) polytype – lonsdaleite. Confocal Raman spectroscopy in combination with AFM was used in an attempt to observe lonsdaleite since Raman peak position of h-dia is downshifted to 1319 cm$^{-1}$ from its usual position at 1332 cm$^{-1}$ (Knight & White, 1989). However, despite very high quality of obtained spectra in a mapping mode no deviations from spectra of conventional cubic diamond were found. This could be explained either by very small relative volume of the eventual h-dia phase and/or by its absence due to another type of atomic arrangement in the twin boundary.

Twinning in diamond is a well known phenomenon. For example, in synthetic diamonds residual stresses between the growth sectors with variable nitrogen concentration may be released by twin formation (Tkach & Vishnevsky, 2004). Observations of deformation twinning in natural diamonds were also reported (Fersmann & Goldschmidt, 1911, Varma, 1970, 1972). Nevertheless it is usually believed that twinning is seldom encountered in natural diamonds and they are normally deformed by slip plane development and dislocation generation and motion. Reliable observations of abundant microtwinning are rare and are usually confined to rare pink and some brown diamonds where macroscopic deformation-related features are clearly visible (Titkov et al., 2012, Gaillou et al., 2010) as well as for heavily deformed crystals (deVries, 1975; Shiryaev et al., 2007). However, the present work shows that microtwinning can be a fairly common phenomenon. Moreover, this approach permits to ascribe unusual lenticular defects reported by Walmsley et al. (1987) to twin-related cracks.

### 3.3. Trigons and related features

The studied surface of the NaCl sample contains several trigons of different sizes and extent. Several interesting properties of the trigon were revealed in the current work. A close look at the interface of the trigon wall with the 111 face shows that at least in some cases the "topmost" 20-40 nm of the trigon wall is steeper than its deeper parts (Fig. 6). The appearance of such step might be related to mechanism of atoms removal during trigon formation.

Namely, the phenomenon described as the dissolution step bunching with formation of so-called negative bunch (Sangwal, 1987 and references therein) may be responsible for this geometry. Presumably, at least in some cases the rate of diamond etching strongly depends on the geometry of the strain fields and an elementary dissolution step sufficiently distant from the stress concentrator may be caught by subsequent (deeper) steps. Such scenario is supported by observation of fairly shallow depressions covering large area around some of the trigons. Similar pattern develop in case of high etch rate which is influenced by impurities in the material. Dislocation-related etch pits with pronounced steps walls may also be produced if the dislocation is decorated by clusters of point defects. Presence of vacancy clusters along deformation-induced dislocations in diamonds was observed by Transmission Electron Microscopy (Bangert et al., 2009).

Walls of the trigons are often decorated by pyramids protruding from the bulk (Fig. 6). The sizes of these features span can differ markedly. They might manifest large-angle grain boundaries resulting from post-growth deformation, but at present the origin of these objects is yet unknown.

### 3.4. Unusual surface features

Figure 7a shows rather unusual mutual arrangement of extended defects: they are misaligned for approx. 11 degrees. Moreover, the phase image of this region (Fig. 7b) shows that the defects indeed possess somewhat different properties, e.g. friction coefficient. One of the most plausible explanations is that the defects are formed by faces with different Miller indices, which are likely characterised by differences in friction coefficient between the AFM cantilever.

Presumably similar defect is shown on figure 7c, illustrating a long trench approx. 1.5 nm deep misoriented relative to predominant features.

### 3.5. Non-crystallographic defects

High stiffness of diamond lattice controls configuration of most extended defects. However, our study revealed existence of interesting surface features without obvious crystallographic control. It is important to emphasise that these features are characterized by *positive* relief.

The first type of such defects is represented by a curious shape of roughly pyramidal hillock (Fig. 8). The origin of this feature is uncertain, but it might consist of several small interacting hillocks in complex mutual arrangement or an impurity precipitate. Another possible explanation is observation of a vertex related to branching dislocations. In general, three slip planes intersect each other in one point only. This anchored vertex cannot perform conservative motion (Kleinert, 1989) and its spatial position is highly localised. Spatial distribution of stresses around such defect can be fairly complex and thus material removal during dissolution might be governed not by general crystallography and be slower than normal etch rate (Seiter, 1977). Another possibility of formation of the positive feature is the impurity segregation on dislocations. The segregation may alter the chemical potential of the local volume, thus altering the dissolution rate (e.g., Sangwal, 1987).

The second type of "non-crystallographic defects" – a comet-like positive features (Fig. 9). In some occurrences (as shown in the figure 9) these defects originate at a linear defect, clearly having a crystallographic control. However, the "comets" may also be unrelated to other defects. A high resolution examination of these "comets" shows existence of quasi-periodic dips along the feature with depth around 1 nm. Though the unique explanation of these defects is yet missing, we believe that they might be related to dislocations intersecting with the crystal face under oblique angle. These dislocations should not be growth-related, but instead are due to stress release around some other defects. The dips might be related to kinks or jogs on the dislocation line.

## 5. Conclusions

In this work results of investigation of surface features on nearly 111 diamond surfaces using Atomic Force Microscopy are presented. AFM clearly has a great potential for investigation of surface features in diamonds, thus complementing more widely used electron and optical microscopy. However, great care in samples preparation and measurement protocol are extremely important.

This work shows that deformation of natural diamond by twining is much more common event than is usually believed. This implies that many diamond crystals were mechanically deformed at temperatures below the Brittle-Ductile transition point. At geologically relevant pressures 1-6 GPa the brittle-to-ductile transition (BDT) occurs at temperatures 900-1100 °C (DeVries, 1975). The majority of previously observed deformation twins (Varma 1970, 1972, DeVries 1975, Gaillou et al., 2010) were macroscopic, often penetrating the whole body of a crystal. Abundant twinning on microscopic scale such observed here and in (Shiryaev et al., 2007, Titkov et al., 2012) is reported less frequently, probably, due to difficulties in observation. The extent of a twin lamella depends mostly on duration and magnitude of external load (e.g., Boyko et al., 1994). Presence of microtwins without obvious large twin lamellae might indicate transient character of external loading or elastic release of twinning. In case of the elastic release the numeric density of remnant microtwins will depend on concentration of pinning defects.

Remarkably, several completely independent analytical techniques (AFM, Small-Angle Scattering, TEM) give indications that at least in some deformed diamonds the extended defects may form fairly regular arrays. The origin of such ordering deserves further investigation, but it might be related to superposition of stress fields from several sources (e.g., dislocations).

**6. Acknowledgements:** We are grateful to professors JP Hirth and VS Boyko for useful discussions of twinning. Mr. Shelkin (NT-MDT) has provided an access to AFM-Raman facility. This work was partially supported by RFBR grant 12-05-00208.

**7. References**

Bangert, U., Barnes, R., Gass, M.H., Bleloch, A.L. & Godfrey, I.S. (2009): Vacancy clusters, dislocations and brown colouration in diamond. *J. Phys.: Condens. Matter.*, **21**, 364208 (6pp).

Boyko, V.S., Garber, R.I., & Kossevich, A.M. (1994): Reversible Crystal Plasticity. AIP ed., New York.

Christian, J.W. & Mahajan, S. (1995): Deformation twinning. *Progress Mater. Sci.*, **39**, 1-157.

DeVries, R.C. (1975): Plastic deformation and "work-hardening" of diamond. *Mater.Res.Bull.,* **10**, 1193-1200.

Gaillou, E., Post, J.E., Bassim, N.D., Zaitsev, A.M., Rose, T., Fries, M.D., Stroud, R.M., Steele, A., Butler, J.E. (2010): Spectroscopic and microscopic characterizations of color lamellae in natural pink diamonds. *Diam. Relat. Mater.,* **19**, 1207–1220.

Fedortchouk, Y., Manghnani, M.H., Hushur, A., Shiryaev, A., Nestola, F. (2011): An atomic force microscopy study of diamond dissolution features: The effect of $H_2O$ and $CO_2$ in the fluid on diamond morphology, *Amer. Miner.*, **96**, 1768–1775.

Fedortchouk, Y., Canil, D., Semenets, E. (2007): Mechanisms of diamond oxidation and their bearing on the fluid composition in kimberlite magmas. *Amer. Miner.*, **92**, 1200-1212.

Fersmann, A, & Goldschmidt, V. (1911): Der Diamant. C. Winter's Universitatsbuchhandlung ed., Heidelberg, Germany.

Hilchie, L. & Fedortchouk, Y. (2009): An experimental study of diamond dissolution in Cl-$H_2O$ systems: implications for mechanisms of diamond oxidation and kimberlitic fluids. *in* Abstracts of Joint Assembly AGU-GAC-MAC, Toronto.

Hirth, J.P., Spanos, G., Hall, M.G., Aaronson, H.I. (1998): Mechanisms for the development of tent-shaped and invariant-plane-strain-type surface reliefs for plates formed during diffusional phase transformations. *Acta Mater.*, **46**, 857-868.

Khokhryakov, A.F. & Pal'yanov, Yu.N. (2007): Revealing of planar defects and partial dislocations in large synthetic diamond crystals by the selective etching. *J. Cryst. Growth*, **306**, 458–464.

Kleinert, H. (1989): Gauge field in condensed matter. Vol II: Stresses and defects. World Scientific ed., Singapore, p. 743-1456.

Knight, D.S. & White, W.B. (1989): Characterization of diamond films by Raman spectroscopy. *J. Mater. Res.*, **4**, 385.

Kosevich, A.M. & Boiko, V.S. (1971): Dislocation theory of elastic twinning of crystals. *Sov. Phys. Usp.*, **14**, 286-342.

Likhachev, V.A., Panin, V.E., Zasimchuk, E.E., et al., (1989): Collective deformation processes and localization of deformation, Kiev, Naukova Dumka, 320. p

Orlov, Y.L. (1963): Morfologiya almaza, Nauka ed., Moscow, 235 p.

Sangwal, K. (1987): Etching of crystals: Theory, Experiment and Applications. North-Holland, Amsterdam.

Seiter, H. (1977): in "Semiconductor Silicon 1977", eds. H. Huff, E. Sirtl; Electrochem. Soc. Proc. Series, p. 187.

Shiryaev, A.A. (2007): Small-angle neutron scattering from diamonds, *J.Appl. Cryst.*, **40**, 116-120.

Shiryaev, A.A., Frost, D.J., Langenhorst, F. (2007): Impurity diffusion and microstructure in diamonds deformed at HPHT conditions. *Diam. Relat. Mater.*, **16**, 503-511.

Shiryaev, A.A. & Boesecke, P. (2012): Small-Angle X-ray and neutron scattering from diamond single crystals. *J. Phys.: Conf. Ser.*, **351**, 012018.

de Theije, F. (2001): Diamond Surfaces - Growth and Etching Mechanisms, PhD thesis, Nijmegen University.

Titkov, S.V., Krivovichev, S.V., Organova, N.I. (2012): Plastic deformation of natural diamonds by twinning: evidence from X-ray diffraction studies. *Miner. Mag.*, **76**, 143-149.

Tkach, V.N, & Vishnevsky, A.S. (2004): Investigation of diamond single crystals of various origin using Kossel method, *in* ''Superhard materials. Synthesis and applications. – Vol. 2,

Structure and properties of superhard materials, methods of investigation'', Alkon, 288 p., Kiev.

Varma, C.K.R. (1970): Evidence of deformation twinning in natural diamond crystals. *J. Phys.Chem. Solids*, **31**, 890-892.

Varma, C.K.R. (1972) Deformation twinning in diamond and identification of the twinning planecrystals. *Scr.Metall.*, **6**, 383-386.

Walmsley, J.C., Lang, A.R., Rooney, M.-L.T., Welbourn, C.M. (1987) Newly observed microscopic planar defects on (111) in natural diamond. *Philos. Mag. Lett.*, **55**, 209-213.

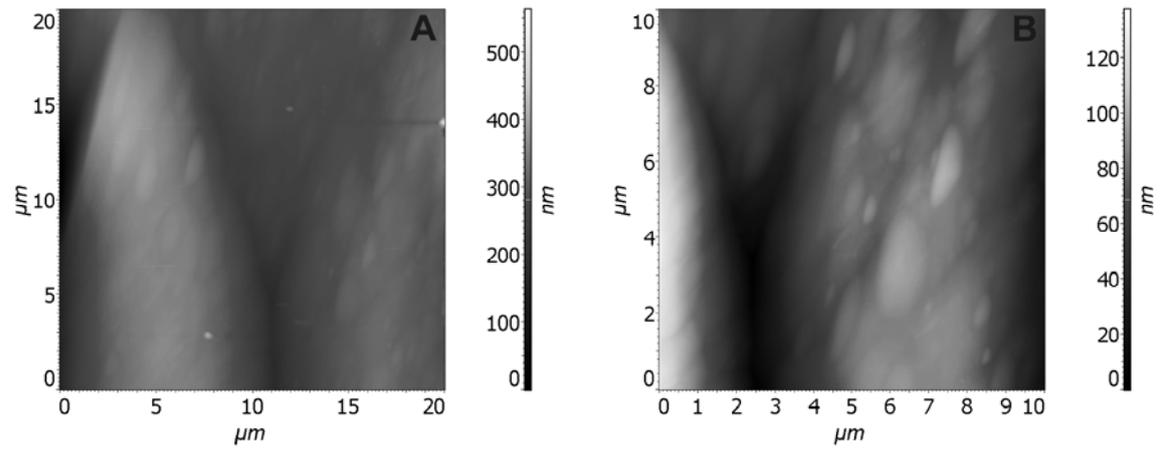

**Figure 1.** Atomic force microscopy images of hillocks of different sizes. A – large scan; B – zoomed part of A showing existence of small asymmetric hillocks.

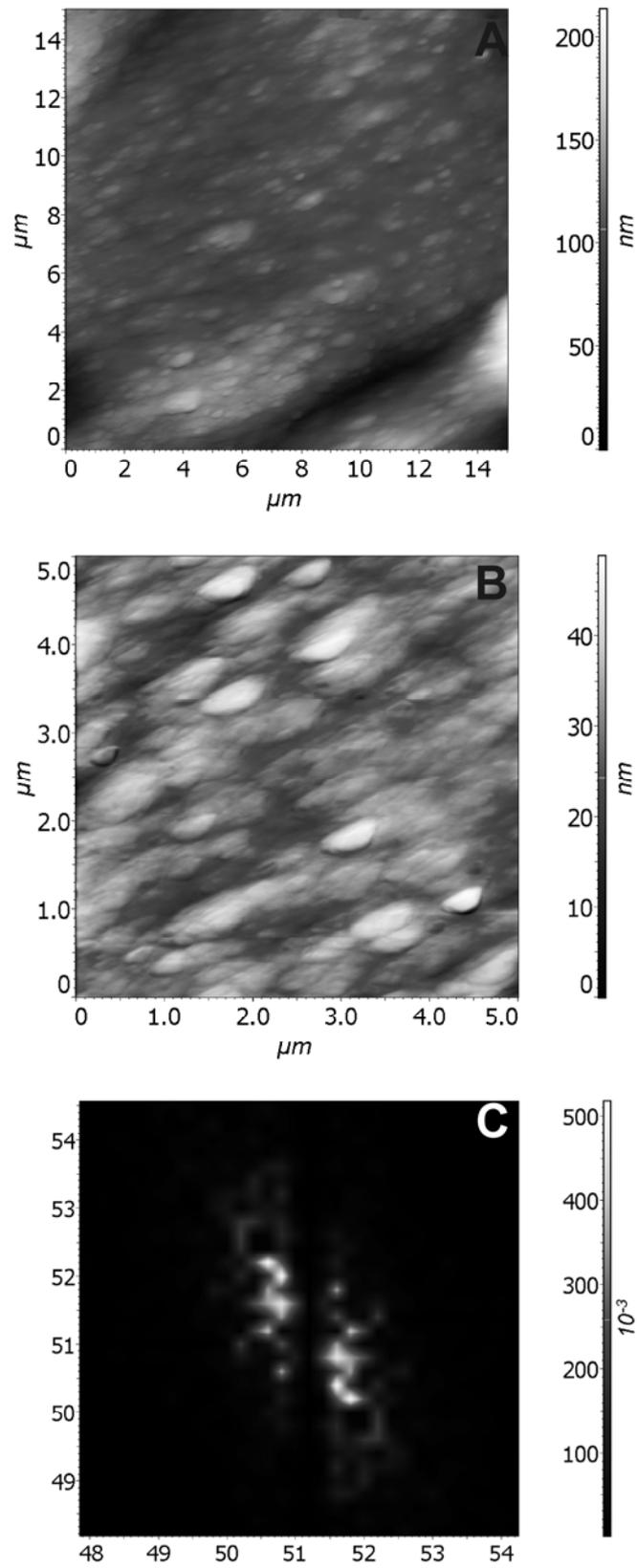

**Figure 2.** Hillocks of different sizes. C - Fourier transform of the image B manifesting existence of ordering in hillocks spatial distribution.

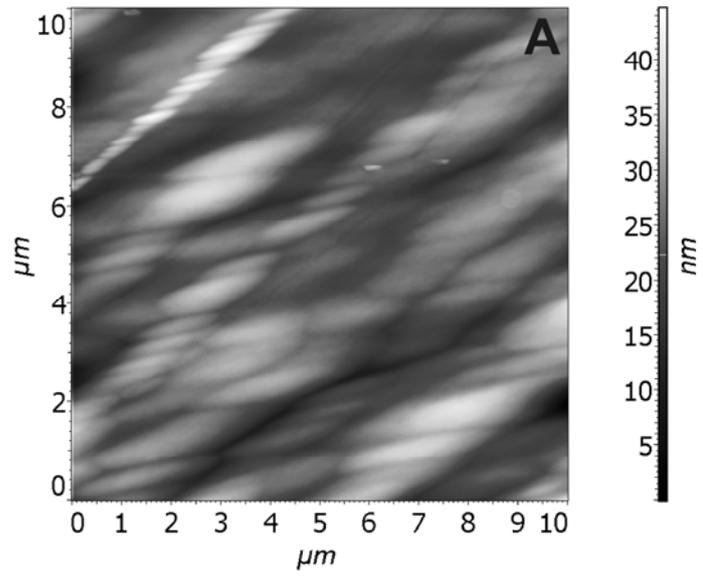

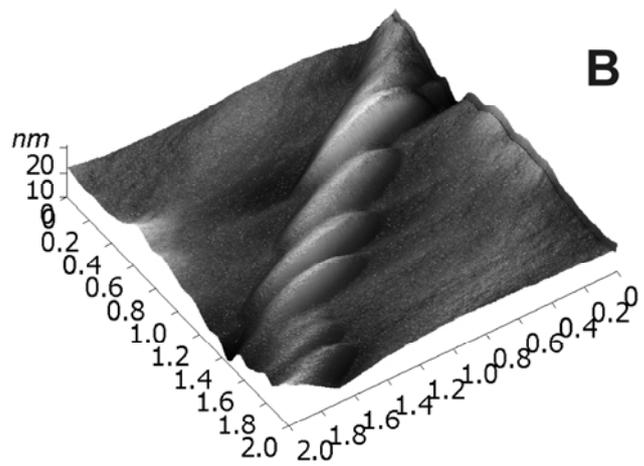

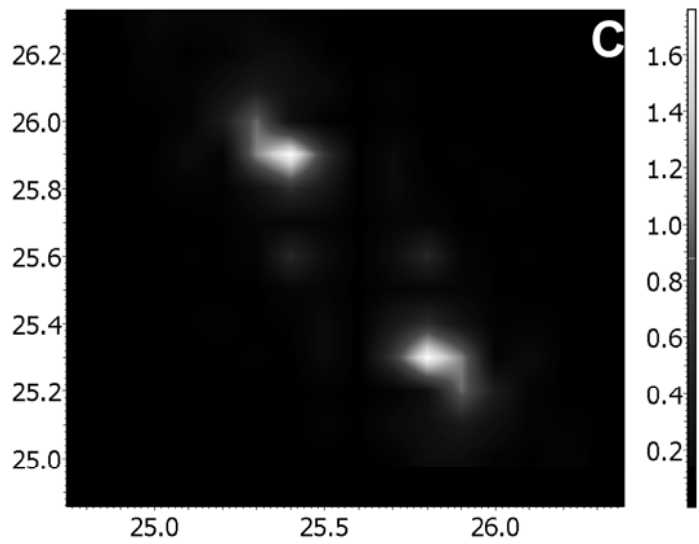

**Figure 3.** Elongated etch hillocks ordered in one direction on surface of one of the diamonds. Figure B shows quasi-3D representation of a small part of the upper left corner of the A.

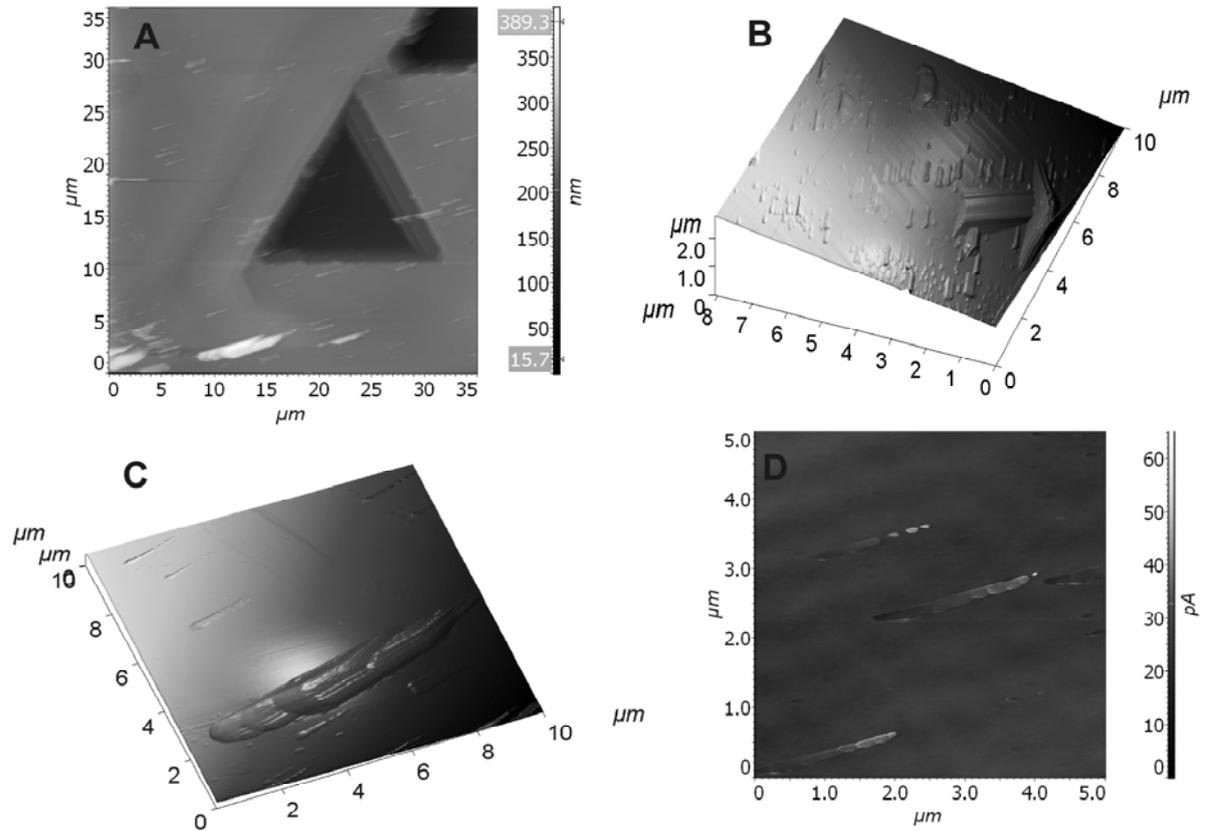

**Figure 4.** Anealastic mechanical twins on the surface of a diamond etched in NaCl aqueous fluid. A – large scale image, numerous bright streaks of variable sizes (the twins) cut the trigon indicating that they are not purely surficial features. B – quasi-3D representation of slightly different spot of the same diamond; the anelastic twins are seen as streaks. In the right and in the upper parts of the image another type of twin – invariant plane strain (ISP) twin (see text). C – higher magnification quasi-3D image of a large anelastic twin. D – a phase image of a family of small twins. Anelastic twins shown here are blunted, indicating presence of defects which have stopped propagation of the twins.

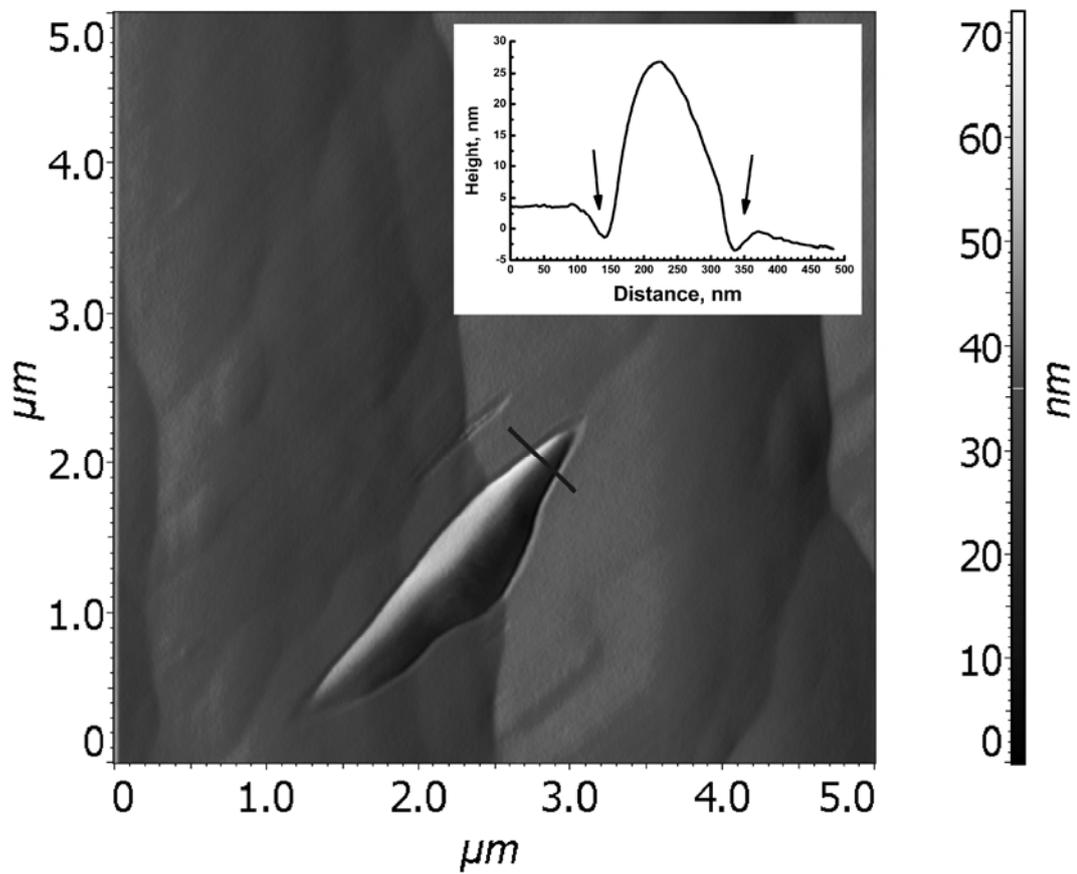

**Figure 5.** "Free" anelastic twins superposed on large scale regular etch bands and hillocks. The inset shows profile across the larger feature (black line). Dips around the twin correspond to twinning dislocations. A small trench in the center corresponds to early stages of another emerging twin.

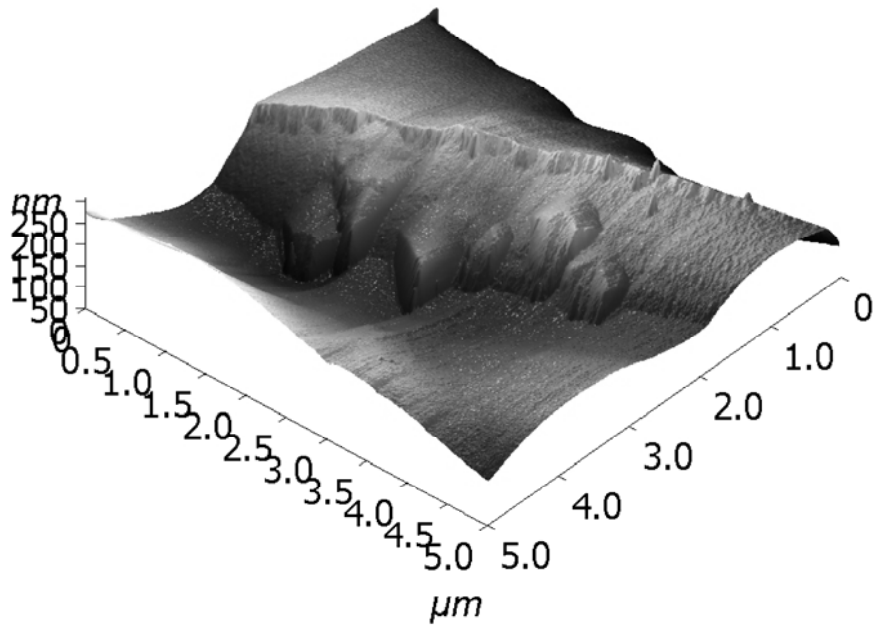

**Figure 6.** A quasi-3D image of unusual features protruding from the wall of a trigon (see also fig. 4a). A steep step in the upper part of the trigon is also seen (see text).

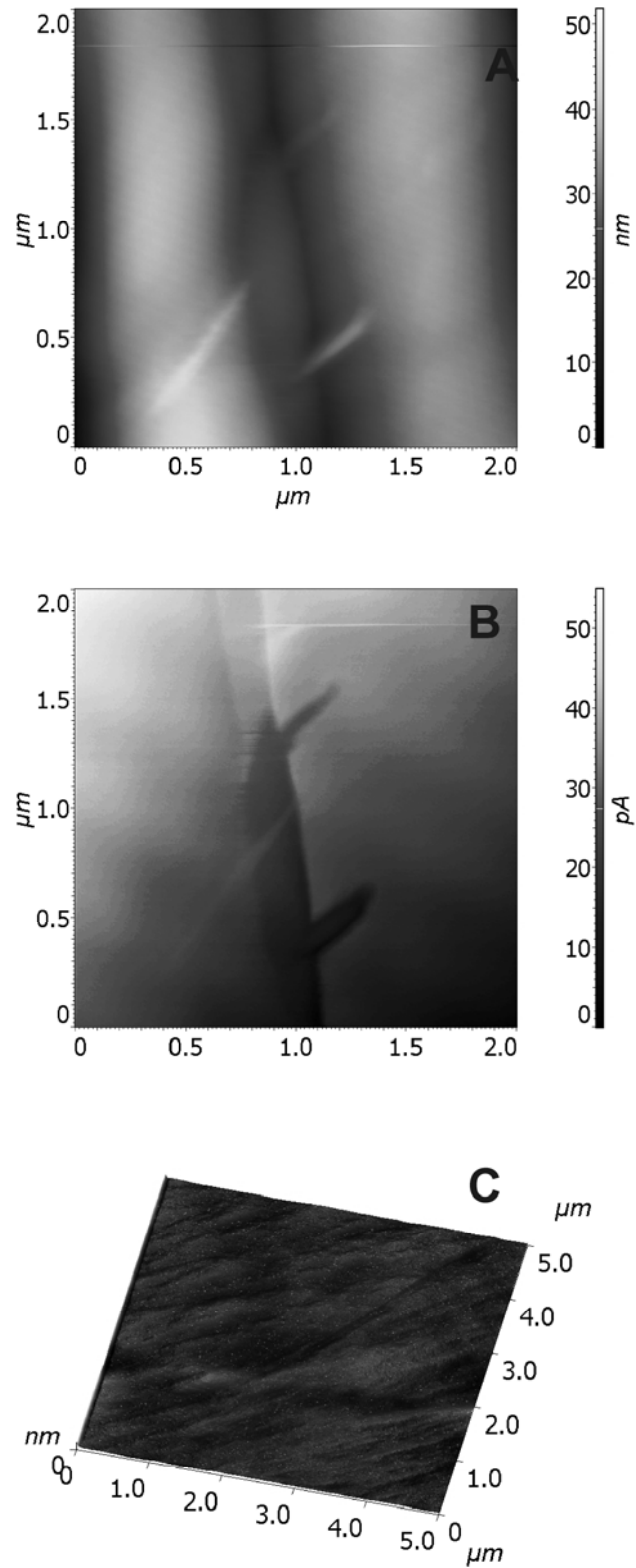

**Figure 7.** Topography (A) and phase (B) image of crystallographically misoriented features. C – quasi-3D image of a shallow trench misoriented relative to other features.

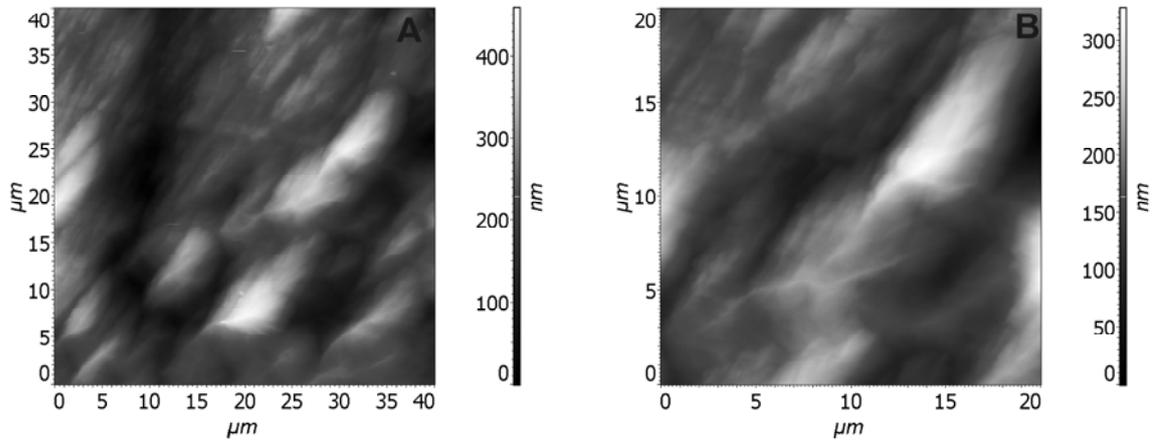

**Figure 8.** Hillocks with sharp vertexes, forming complex relief.

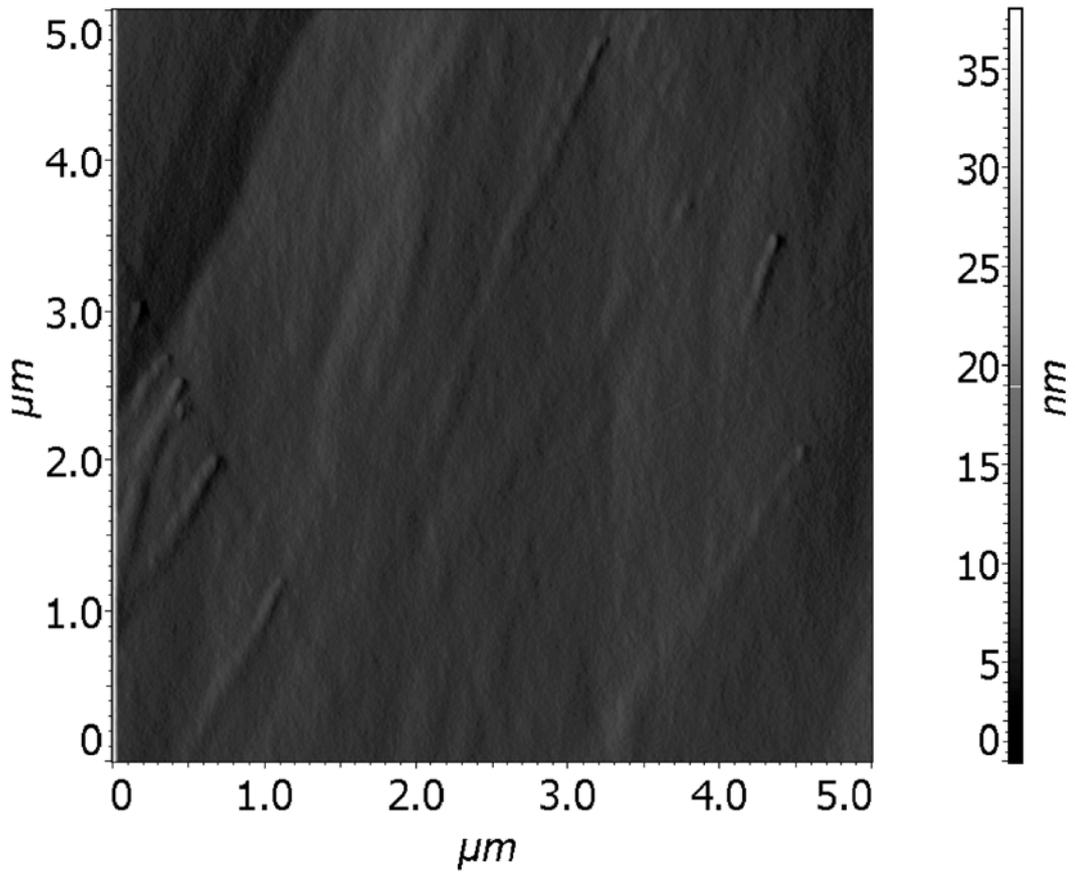

**Figure 9.** Comet-like positive relief features. The original image was enhanced using Sobel filter. Note that those in the left lower corner originate at some linear defect. The features probably correspond to dislocations subparallel to the surface.